\documentclass[aps,pra,floatfix,twocolumn]{revtex4}
\usepackage{amsmath}
\usepackage{amssymb}
\usepackage{mathrsfs}
\usepackage{graphicx}
\usepackage[francais]{babel}
\newcommand{\ipr}{Institut de Physique de Rennes, Universit\'e Rennes I - CNRS UMR 6251,Campus de Beaulieu, 35042 Rennes Cedex, France}
\begin{document}
\title{Measuring the universal synchronization properties of coupled oscillators across the Hopf instability}
\author{M. Romanelli, L. Wang, M. Brunel, and M. Vallet}
\affiliation{\ipr}
\begin{abstract}
When a driven oscillator loses phase-locking to a master oscillator via a Hopf bifurcation, it enters a bounded-phase regime in which its average frequency is still equal to the master frequency, but its phase displays temporal oscillations. Here we characterize these two synchronization regimes in a laser experiment, by measuring the spectrum of the phase fluctuations across the bifurcation. We find experimentally, and confirm numerically, that the low frequency phase noise of the driven oscillator is strongly suppressed in both regimes in the same way. Thus the long-term phase stability of the master oscillator is transferred to the driven one, even in the absence of phase-locking. The numerical study of a generic, minimal model suggests that such behavior is universal for any periodically driven oscillator near a Hopf bifurcation point.


\end{abstract}
\maketitle
\section{Introduction}
Synchronization, i.e. the tendency of coupled oscillators to oscillate at the same frequency, is recognized as ``one of the most pervasive drives in universe''~\cite{Strogatz2003,Pikovsky2003}. Countless examples of this fact have been reported, from the synchronous blinking of fireflies to the coherent oscillation of coupled Josephson junctions~\cite{Wiesenfeld1996} or of chemical micro-oscillators~\cite{Toiya2010}. Synchronization plays also a role in memory processes~\cite{Fell2011} and in the complex networks arising, for instance, in neurosciences or in economics~\cite{Arenas2008}. Given that synchronization occurs in so many different contexts, the study of a specific system can acquire a broader significance by helping to identify features that are universal, i.e. common to a large class of different systems displaying synchronization.\newline
\indent
In laser physics, issues related to synchronization are typically encountered in optically injected lasers. Optical injection has been studied for a long time, but it is still a subject of interest today, both for the great complexity and variety of the possible dynamics~\cite{Wieczorek2005, Erneux2010}, and for its importance for applications such as high-speed modulation~\cite{Grillot2009} or microwave photonics~\cite{Hwang2013}. In particular, recently a synchronization regime of frequency locking without phase locking has been experimentally isolated in injected laser oscillators~\cite{Kelleher2010,Thevenin}. In this regime, the relative phase of two interacting oscillators is not constant in time, yet its variations are bounded, so that the systems keep oscillating at the same average frequency. It is quite obvious that this regime differs from the more common situation in which the oscillators are phase-locked, but how should one characterize and quantify this difference? In the present paper, we propose to answer this question by studying experimentally and numerically a driven, optically-based radiofrequency (RF) oscillator. Furthermore, the bounded-phase behavior, predicted long ago in driven laser systems~\cite{Braza1990, Solari1994, Strogatz1998} and more recently also in quantum dot lasers~\cite{Ludge2012, Ludge2013}, is by no means specific to lasers, since it occurs also in models of circadian~\cite{Kronauer1982} or coupled van der Pol~\cite{Chakraborty1988} oscillators. Actually, bounded-phase phenomena are extremely generic, since they are associated to a Hopf bifurcation~\cite{Kelleher2012}, and are to be expected when amplitude effects cannot be ignored in the dynamics of coupled oscillators~\cite{Aronson1990}. Thus, it can be expected that our study of a driven opto-RF oscillator will be relevant for coupled oscillators in general. Indeed, we will show in the following that our system provides an example of a generic, universal behavior of driven oscillators close to a Hopf bifurcation point.
\section{Experimental  setup and results}
\subsection{Experimental setup}
\begin{figure}[h]
\centering\includegraphics[width=8cm]{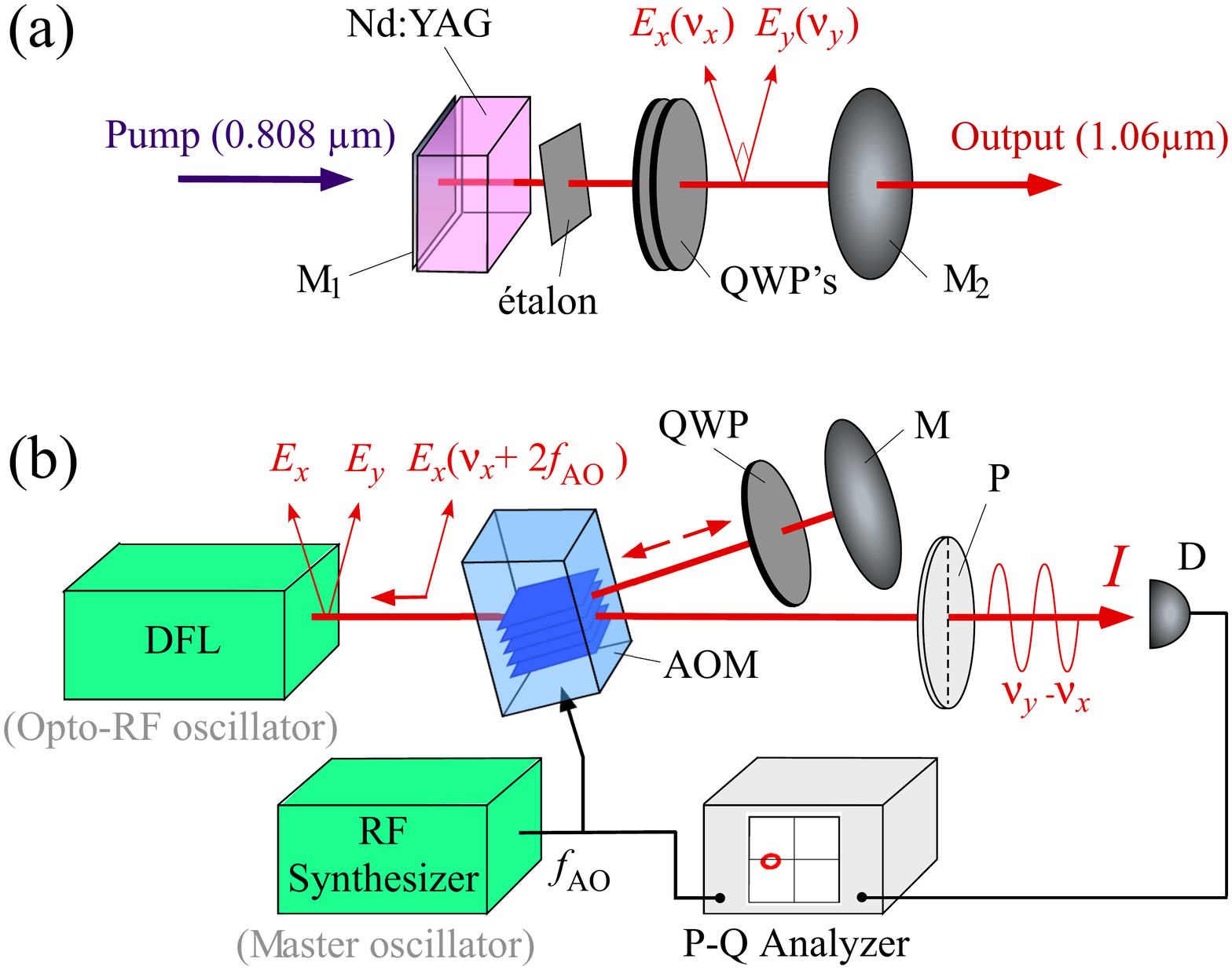} \caption{(a) Dual-Frequency Laser. $M_{1,2}$: cavity mirrors. QWP: quarter-wave plate. (b) Experimental setup that allows synchronizing the DFL beat-note frequency to a RF synthesizer. DFL: Dual-frequency laser (slave oscillator). M: feedback mirror. QWP: quarter-wave plate. AOM: acousto-optic modulator. P: polarizer. D: detector. P-Q analyzer: digital vector signal analyzer, permitting to measure the quadratures of \textit{I} with respect to the reference signal delivered by the RF synthesizer.} \label{fig1}
\end{figure}

The experimental setup (see Fig.~\ref{fig1}) is similar to the one described in~\cite{Thevenin2012}. 
We build an optically-based RF oscillator, and drive it with a stable RF synthesizer, that is coupled to the slave oscillator by optical frequency-shifted feedback.
The core of the experiment is a diode-pumped, solid-state dual-frequency laser (DFL)~\cite{Brunel2004}, represented in Fig.~\ref{fig1}(a).
The laser cavity, of length L = 75 mm, is closed on one side by a high-reflection plane mirror, coated on the 5-mm long Nd:YAG active medium, and on the other side by a concave mirror (radius of curvature of 100 mm, intensity transmission of 1\% at the lasing wavelength $\lambda$ = 1064 nm). The active medium is pumped by a laser diode emitting at 808 nm. A 1 mm-thick silica \'etalon ensures single longitudinal mode oscillation. Two eigenmodes $E_x$ and $E_y$, polarized along $\hat{x}$ and $\hat{y}$, with eigenfrequencies $\nu_x$ and $\nu_y$ respectively, oscillate simultaneously. An intracavity birefringent element (here two quarter-wave plates QWPs) induces a frequency difference, finely tunable from 0 to $c/4L$ = 1 GHz by rotating one QWP with respect to the other~\cite{Brunel1997}.  The typical output power of the two-frequency laser is 10 mW when pumped with 500 mW.
When the laser output is detected by a photodiode after a polarizer at 45$^\circ$, an electrical signal oscillating at the frequency difference $\Delta \nu_0= \nu_y - \nu_x$ is obtained. The DFL can thus be seen as an opto-RF oscillator. \newline 
\indent
In order to lock this oscillator to an external reference signal, we use optical frequency-shifted feedback~\cite{Kervevan} (Fig.~\ref{fig1}(b)). The feedback cavity contains an acousto-optic modulator (AOM), driven by a stable RF synthesizer, which provides an external phase reference. Next, a quarter-wave plate at 45$^\circ$ followed by a mirror flips the $\hat{x}$ and $\hat{y}$ polarizations, and finally the laser beam is reinjected in the laser cavity after crossing again the AOM. As a result, a $\hat{x}$-polarized field oscillating at the frequency $\nu_y+2 f_{AO}$ and a $\hat{y}$-polarized field oscillating at the frequency $\nu_x+2 f_{AO}$ are reinjected in the laser. We choose the value of $2 f_{AO}$ so that $\nu_x+2 f_{AO}$ is close to $\nu_y$. Under suitable feedback conditions, $\nu_y$ locks to the injected beam frequency $\nu_x+2 f_{AO}$, i.e. the frequency difference $\nu_y -\nu_x$ locks to $2 f_{AO}$. We note that the optical reinjection has no direct effect on $E_x$, because the frequency difference between $\nu_x$ and $\nu_y+2 f_{AO}$ is too large. For the same reason, multiple round trips in the feedback cavity have no effect on the dynamics. The laser output is detected with a photodiode (3 GHz analogic bandwidth) after a crossed polarizer, thus providing an electrical signal proportional to $I=|E_x+E_y|^2$. The signal is then analyzed with an electrical spectrum analyzer, a digital P-Q signal analyzer and an oscilloscope.
\subsection{Experimental results}
\begin{figure}[h]
\centering\includegraphics[width=0.5\textwidth]{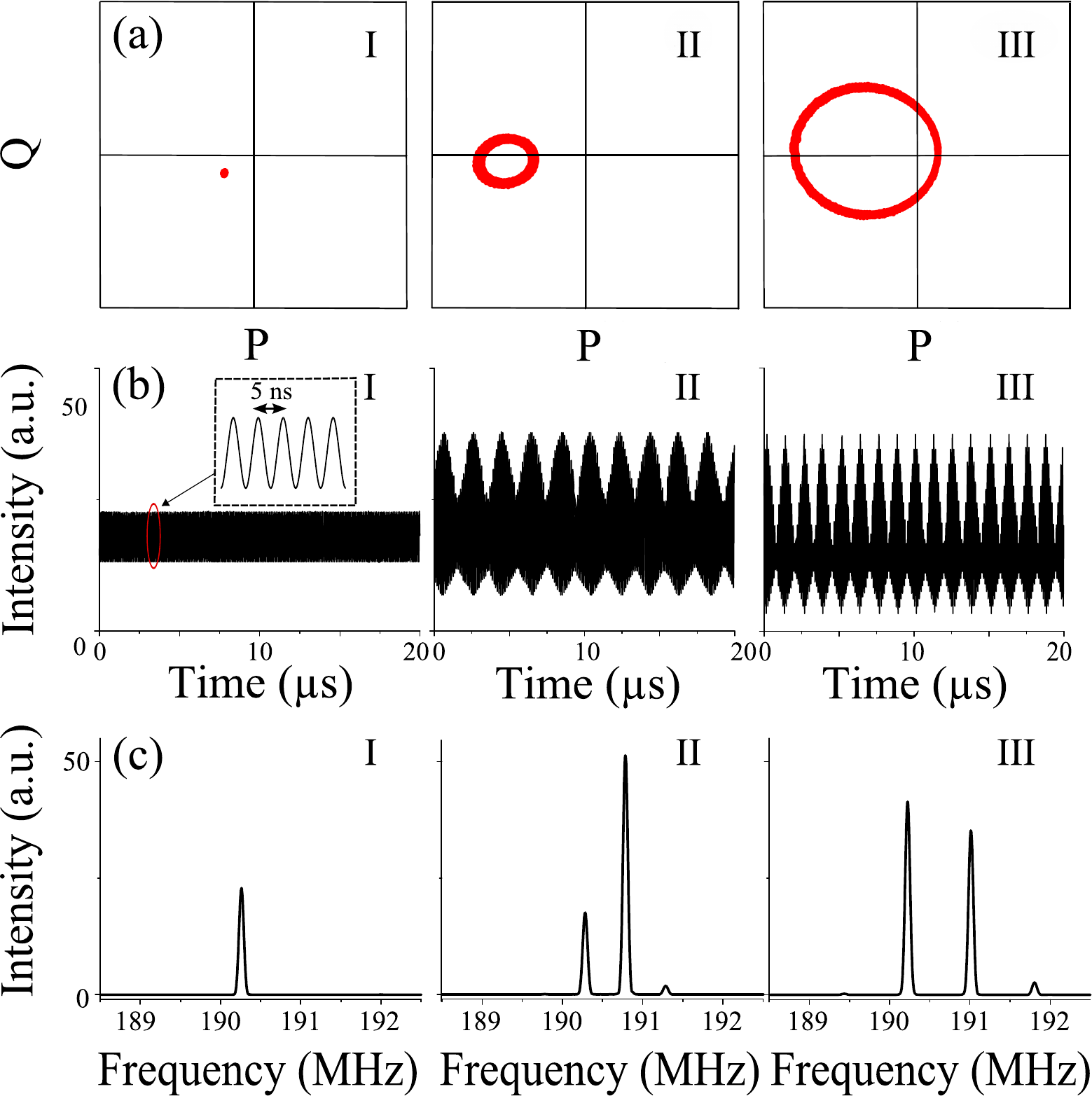} \caption{(a) Experimental phasor plots of the output signal $I=|E_x+E_y|^2$, in the reference frame of the local oscillator at $2f_{AO}$. Each plot contains 2000 points, recorded over 100 $\mu$s. (I) Phase-locking regime. (II) Bounded-phase regime. (III) Unbounded-phase regime. (b) Corresponding experimental time series. (c) Power spectra.}
\label{fig2}
\end{figure}
An electrical signal, oscillating in the absence of feedback at the frequency difference $\Delta \nu_0= \nu_y - \nu_x$ around 200 MHz, is obtained by detecting the interference signal $I=|E_x+E_y|^2$  between the laser fields on a photodiode after a polarizer. The whole system can thus be seen as an opto-RF oscillator, driven by the RF synthesizer (master oscillator) when the optical feedback is present.
The beat-note signal $I$ can be visualized at each instant in time as a vector having two components $P = I_0(t) \cos \phi(t)$ and $Q = I_0(t) \sin \phi(t)$ in a reference frame rotating at the master signal frequency $2f_{AO}$. The phase $\phi(t)$ is defined as $\phi(t) = \phi_y(t)-\phi_x(t)-2 \pi 2 f_{AO} t$, where $\phi_{x,y}$ are the optical phases of the two electric fields $E_x$ and $E_y$. For small enough detuning $\Delta \nu= \Delta \nu_0 - 2 f_{AO}$ between the slave and the master oscillator, phase-locking occurs, and the corresponding experimental phasor plot consists of just a fixed point (Fig.~\ref{fig2}(a), regime I). If $\Delta \nu$ is increased above the Adler frequency $f_A$, the fixed point solution loses stability through a Hopf bifurcation. The point representing $I$ then rotates at the frequency $\Delta \nu$ on a limit cycle (Fig.~\ref{fig2}(a), regime II). The relative phase exhibits bounded, periodic temporal oscillations, whose amplitude increases with $\Delta \nu$. Finally, when $\Delta \nu$ is bigger than a second characteristic frequency $f_B$, the origin of the plane lies inside the limit cycle and the phase variations are not bounded anymore (Fig.~\ref{fig2}(a), regime III).
In the time and frequency domains (Fig.~\ref{fig2}(b-c)), phase-locking (I) appears as pure sinusoidal oscillation, and the corresponding power spectrum consists of a single, sharp Fourier peak. On the contrary, in the bounded (II) and in the unbounded-phase regimes (III), the power spectrum shows two peaks at the natural frequencies $2 f_{AO}$ and $\Delta \nu_0$ of the uncoupled master and slave oscillator respectively, and the time series displays a beating, i.e. a slow modulation of the amplitude of the oscillations.
From Fig.~\ref{fig2}, the question arises about the exact status of the bounded-phase regime, occurring when $f_A < \Delta \nu < f_B$. On the one hand, since the relative phase is bounded, the average frequency of the oscillators is the same, and they can be considered as synchronized. On the other hand, the bounded-phase regime appears very similar to the unbounded-phase regime. In particular, the power spectrum displays two separated peaks at the natural frequencies of the oscillators, as it happens when synchronization is absent. Furthermore, while a qualitative change of the dynamics occurs when passing from phase locking to the bounded phase through a Hopf bifurcation, in the $\{$P, Q$\}$ plane in Fig.~\ref{fig2}(a) the bounded-phase and unbounded-phase regimes appear topologically equivalent and no bifurcation is seen between them~\cite{Wieczorek2005}.\newline
\indent
To investigate quantitatively such questions, we propose to use the power spectral density (PSD), also called phase spectrum, of the relative phase between the oscillators as a meaningful measure of synchronization. The PSD measures the strength of the phase fluctuations of an oscillator at all frequencies, and thus characterizes its quality in a complete and rigorous way.
Precisely, the PSD is defined as follows~\cite{Rubiola2008, IEEE2008}. 
For a stochastic, ergodic process $\phi (t)$, the autocorrelation function is:
\begin{equation}
C(\tau)=\left\langle \phi (t) \phi (t+\tau)\right\rangle = \lim_{T \to \infty}\frac{1}{T}\int_{-T/2}^{T/2} \phi (t) \phi (t+\tau) dt.
\end{equation}
Brackets indicate the ensemble average. The single-sideband power spectral density of the process is defined as
\begin{equation}
PSD(f)=2 \theta (f) \int_{-\infty}^{+\infty} C(\tau) \exp{(-i 2 \pi f \tau)} d\tau,
\end{equation}
where $\theta (f)$  is the Heaviside function. Indeed, the negative frequency components can be discarded because $C(\tau)$ is a real, even function. $PSD(f)$ is frequently expressed in decibels as 10 $\log \left[PSD(f)\right]$, and its units are dBrad$^2$/Hz. This is the quantity we have measured, and indicated as $S_\phi (f) = 10 \log \left[PSD(f)\right]$. In practice, the above definitions are not directly used for the computation of $S_\phi (f)$. One rather uses the equality $PSD(f) \, df = 2 \theta(f) \left\langle|d\Phi(f)|^2\right\rangle$, and estimates $|d\Phi(f)|^2$ by taking the square modulus of the Fast-Fourier Transform (FFT) of a sampled time series of $\phi (t)$.
\begin{figure}[h]
\centering\includegraphics[width=0.5\textwidth]{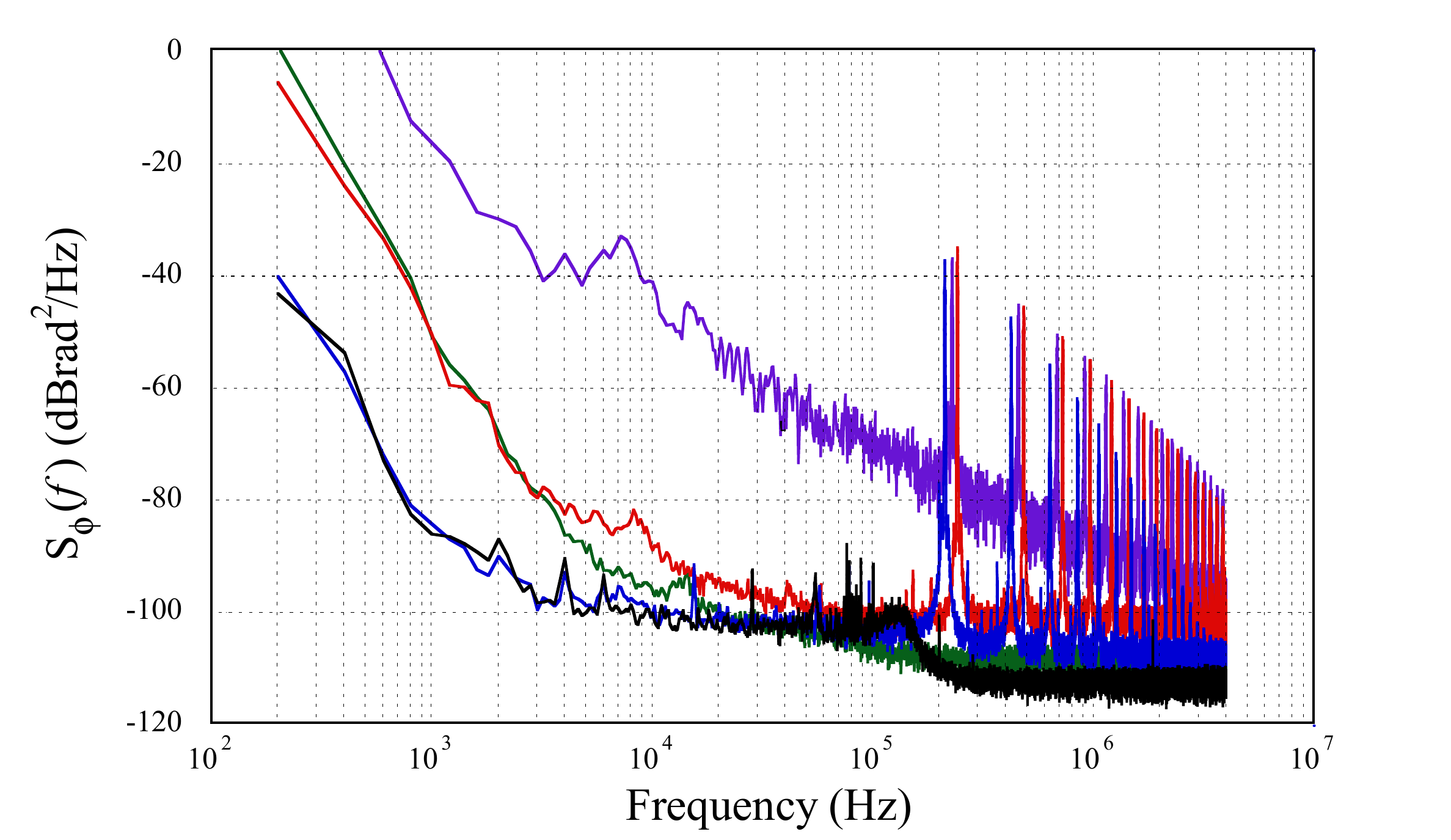} \caption{Measured phase spectra $S_\phi(f)$ for different values of the detuning $\Delta \nu$. \newline Black: $\Delta \nu$ = 60 kHz, phase locking regime. Blue: $\Delta \nu$ = 220 kHz, bounded-phase regime. Violet: $\Delta \nu$ =  240 kHz $\simeq f_B$. Red: $\Delta \nu$ = 260 kHz, phase-drifting regime. Green : free-running laser (no feedback from the external cavity).}
\label{fig4}
\end{figure}
In our setup, the relative phase $\phi(t)$ can be extracted from the time series of the two quadratures $P(t)$ and $Q(t)$.
The spectrum $S_\phi(f)$ of $\phi(t)$ is then computed. Note that, when the phase is unbounded, the average slope of the drifting phase has to be substracted in order to get a stationary, ergodic process. The experimental spectra are shown in Fig.~\ref{fig4}. One sees that, in the phase-locking regime, the phase fluctuations of the free running laser are strongly suppressed at low frequency. For instance, at 1 kHz from the carrier one has a noise reduction of more than 30 dB with respect to the free-running case. The phase of the slave oscillator is locked to the stable master, which ensures the long-term phase stability. The absolute value of the phase noise that we obtain, $-$100 dBrad$^2$/Hz at 10 kHz from the carrier, is comparable to what is obtained with active stabilization techniques using electronic servo-loops~\cite{Brunel2004b}.
\begin{figure}[h]
\centering\includegraphics[width=0.5\textwidth]{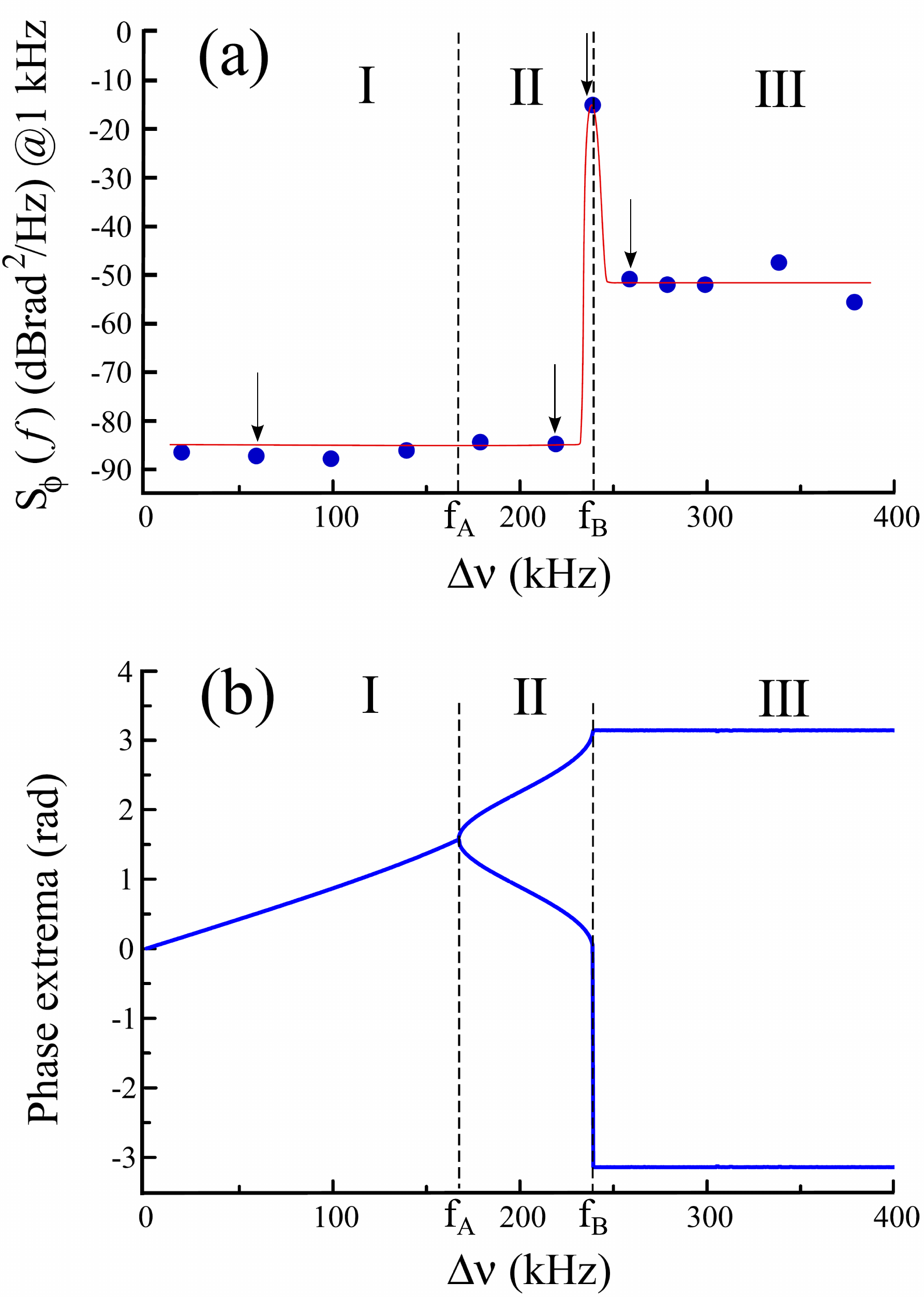} \caption{(a) Measured phase spectra $S_\phi(f)$ at 1 kHz from the carrier frequency, as a function of the detuning $\Delta \nu$. The arrows correspond to the spectra of Fig.~\ref{fig4} (except for the free-running laser situation, not represented here). The red line is an eye-guide. (b) A phase bifurcation diagram, calculated using the laser model and the measured values of $f_A$ and $f_B$, is shown for comparison.}
\label{fig5}
\end{figure}
In the bounded-phase regime, it can be seen that the long-term phase stability is exactly the same as in the phase-locking regime. The low frequency part of the spectrum can hardly be distinguished in the two cases. In this case too, despite the occurrence of fast temporal oscillations, the average value of the slave oscillator phase is locked to the stable master phase, and low frequency drifts and fluctuations are removed with respect to the free running laser. The only differences in the spectra arise at high frequency. Indeed, in the bounded-phase regime one has a narrow peak at the detuning frequency $\Delta \nu$, and its harmonics. These peaks are the coherent signature of the Hopf bifurcation, and correspond to the onset of the deterministic oscillation of the phase beyond the bifurcation point $f_A$. At the transition between the bounded and unbounded-phase regime, for $\Delta \nu \simeq f_B$, there is an enormous increase of the phase noise. This can be understood by observing in Fig.~\ref{fig2}(a) that the experimental limit cycle has a finite thickness, because of uncontrolled fluctuations of the parameters and of the setup. Therefore, when $\Delta \nu$ approaches $f_B$, erratic phase slips occur in a random fashion, driven by the noise present in the system. These stochastic, abrupt 2$\pi$ phase jumps close to the point $f_B$ explain the huge amount of phase noise. Finally, the spectrum of the unbounded-phase regime corresponds essentially to the one of the free-running opto-RF oscillator, apart from the high-frequency peaks.
Figure~\ref{fig5}(a) reports the PSD value at 1 kHz from the carrier as a function of $\Delta \nu$. It is striking to compare this plot to a numerically calculated bifurcation diagram (Fig.~\ref{fig5}(b)). Indeed, when looking at the long-term phase stability of the oscillator, the occurrence of a Hopf bifurcation at the point $f_A$ is completely undetectable. Conversely, at the point $f_B$ a sharp transition occurs as a signature of the unlocking of the slave oscillator. The conclusion that can be drawn is that the relevant boundary for the synchronization range is not the Adler frequency $f_A$, but the bounded-phase frequency $f_B$. From the viewpoint of dynamical systems, phase-locking and bounded-phase are two qualitatively different regimes, separated by a bifurcation; yet this distinction turns out to be of little relevance if one is concerned with synchronization.\newline
\indent
At this point, one may ask whether the bounded/unbounded-phase transition is a bifurcation or not. Mathematically, the answer actually depends on the choice of variables~\cite{Wieczorek2005}. If the beat-note signal is described using the amplitude $I_0(t)$ and the phase $\phi(t)$, then its phase space is the surface of a half-cylinder. Bounded-phase and unbounded-phase regimes correspond to limit cycles that do not wrap the half-cylinder (librations) or that wrap it (rotations) respectively, and are not topologically equivalent. On the contrary, if the signal is described using the quadratures $P(t)$ and $Q(t)$, the two regimes are topologically equivalent, as we have seen. This difference stems from the fact that the polar coordinates $I_0(t)$ and $\phi(t)$ are not homeomorphic with the planar coordinates $P(t)$ and $Q(t)$. So, mathematically the nature of the bounded/unbounded-phase transition is, to some extent, a matter of choice, and it is worth noticing that, in a theoretical bifurcation analysis, it looks more convenient to work with $P(t)$ and $Q(t)$, so that no bifurcation appears at $f_B$~\cite{Wieczorek2005}. However, our results show clearly that the physics of the system is not ambiguous: a sharp transition occurs at the point $f_B$, namely the transition from synchronization to desynchronization.
\section{Theory: laser and generic oscillator models}
\begin{figure}[h]
\centering\includegraphics[width=0.5\textwidth]{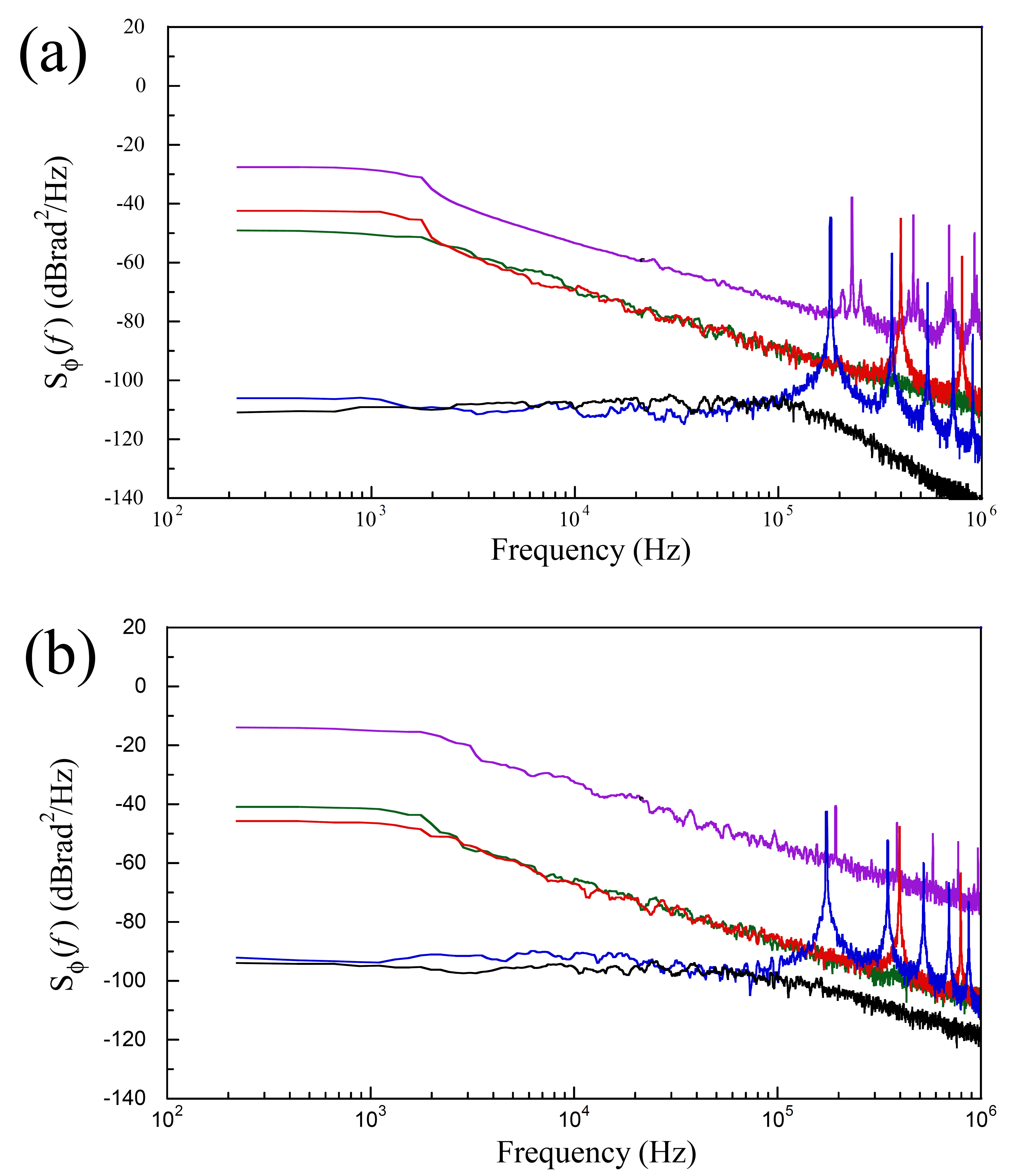} \caption{(a) Simulated phase spectra $S_\phi(f)$ using the laser model. Black: $\Delta \nu$ = 100 kHz, phase locking regime. Blue: $\Delta \nu$ = 180 kHz, bounded-phase regime. Violet: $\Delta \nu$ = 230 kHz $\simeq f_B$. Red: $\Delta \nu$ = 400 kHz, phase-drifting regime. Green: free-running laser (no feedback from the external cavity). The values of the parameters used in the simulations are: $\beta$ = 0.6, $f_A$ = 160 kHz, $f_R$ = 70 kHz, $\gamma = f_A/f_R$, $\epsilon$ = 0.97 10$^{-2}$, $\eta$ = 1.2.
(b) Simulated phase spectra $S_\phi(f)$ for the generic model. Black: $e$ = 1.7, $\Delta$ = 1.4, phase locking regime. Blue: $e$ = 1.7, $\Delta$ = 2.6, bounded-phase regime. Violet: $e$ = 1.7, $\Delta$ = 2.9 $\simeq f_B$. Red: $e$ = 1.7, $\Delta$ = 5.7, phase-drifting regime. Green: $e$ = 0, no forcing.}
\label{fig6}
\end{figure}

We compare the experimental observation with the following laser model~\cite{Thevenin2012}:
\begin{eqnarray}
\frac{de_x}{ds} &=& \frac{\left( m_x + \beta m_y \right)}{1+\beta} \frac{e_x}{2}, \label{norm_e_x} \\
\frac{de_y}{ds} &=& \frac{\left( m_y + \beta m_x \right)}{1+\beta} \frac{e_y}{2} + i \Delta \, e_y + \gamma e_x,\label{norm_e_y}\\
\frac{dm_{x,y}}{ds} &=& 1 - \left(|e_{x,y}|^2 + \beta |e_{y,x}|^2\right)+ \label{norm_m}\\ \nonumber
&-& \epsilon \, m_{x,y}\left[1 + (\eta - 1)(|e_{x,y}|^2 + \beta |e_{y,x}|^2)\right],
\end{eqnarray}
where $e_{x,y}$ are the normalized amplitudes of the optical modes, $ m_{x,y}$ describe the population inversions, $\eta$ is the pump parameter, $\beta$ accounts for the cross saturation in the active medium, $\gamma$ is the feedback strength, $\epsilon$ is the inversion lifetime, $\Delta = \Delta \nu/f_R= (\Delta \nu_0-2f_{AO})/f_R$ is the detuning, and $f_R$ is the relaxation oscillation frequency. The scaled time $s$ is related to the physical time $t$ by $s= 2\pi f_R \, t$. 
The normalization has been chosen so to remove stiffness from the class-B laser equations~\cite{Erneux2010}. For more details on the model, and for the correspondence between the normalized and the physical quantities, see~\cite{Thevenin2012}.
We have injected some noise in the system by replacing the pump parameter $\eta$ with $\eta(1+0.05\xi(s))$, where $\xi(s)$ is a $\delta$-correlated, normally distributed stochastic process. We notice that pump fluctuations induce fluctuations of $f_R$, and thus of the normalized quantities $\Delta$ and $\gamma$. This fact must be taken into account when integrating the laser equations.
The PSDs calculated from numerical integration of the laser model are reported in Fig.~\ref{fig6}(a). As in the experiment, it is observed that optical feedback is effective in reducing the low frequency phase fluctuations, both when the phase is locked and when it is bounded. In both cases, the opto-RF oscillator is thus synchronized to the master, and the long-term phase stability is not reduced by the onset of self-sustained oscillations. The huge increase of phase noise around the point $f_B$ is also reproduced, as well as the fact that, in the unbounded-phase regime, the phase noise is essentially equal to that of the free-running oscillator.
One can ask to which extent these findings are specific to our system. In order to answer this question, we have numerically investigated the following minimal model~\cite{Pikovsky2003}:
\begin{equation}
\frac{dy}{ds} = (1-i\Delta)y-(1+i\alpha)|y|^2y+ e.\label{min}
\end{equation}
$y$ represents the complex amplitude of a periodically driven oscillator, $\Delta$ is the detuning between the master and the slave, $\alpha$ accounts for the possible dependence of the oscillator's phase on the amplitude of the oscillations, and $e$ is the strength of the drive. Equation~\ref{min} contains just a linear term accounting for the growth of oscillations, a saturating nonlinearity, and an external forcing, and is a universal equation describing any periodically driven oscillator close to a supercritical Hopf bifurcation point~\cite{Pikovsky2003}. We have introduced some noise by replacing $\Delta$ with $\Delta \left[1+0.033\,\xi (s)\right]$. 
In the following, we have taken $\alpha$ = 0 because this seems more appropriate for comparison with a solid-state laser. However, some numerical simulations, not shown here, indicate that this parameter does not influence phase noise, when $\alpha^2 < 1/3$. For higher values of $\alpha$, the bifurcation structure of the generic model changes qualitatively~\cite{Pikovsky2003}, and we have not explored this case. 
The PSDs of the phase of $y$ are plotted in Fig.~\ref{fig6}(b), for different values of $\Delta$. 
The agreement of the spectra in Fig.~\ref{fig6}(b) with the laser model is striking. All the observed regimes are reproduced and appear thus to be generic for any driven oscillator undergoing a Hopf bifurcation.
\section{Conclusion}
In conclusion, we have characterized the synchronization of an opto-RF oscillator to a reference oscillator by measuring the noise spectrum of the relative phase, and shown that the long-term stability of the master oscillator is transferred to the driven oscillator even when the latter loses phase-locking and enters the bounded-phase regime. The synchronization range thus extends up to the edge of the bounded-phase regime, and can be substantially larger than the phase-locking range (of a factor $\sqrt{2}$ typically~\cite{Strogatz1998}). A generic, minimal model gives strong evidence that the robustness against a Hopf instability is a universal feature of synchronizing systems.
\newline
\indent
Given the ubiquitous occurrence of synchronization and the genericity of the mechanism discussed here, the present results can be relevant for many areas. In a broader context, recently synchronization phenomena have been highlighted in optomechanical~\cite{Marquardt2011} or micromechanical~\cite{Agrawal2013} oscillators. Most of the recently observed results could be explained in the framework of pure phase equations such as the Adler model, which is the case when the coupling is too weak to affect the amplitude. However, as soon as this condition is not fullfilled, Hopf dynamics is to be expected, therefore we believe that the results presented here will soon prove relevant also for these research  fields. One can also expect the present results to have an impact in the context of delay-coupled systems at large~\cite{Soriano}.
In general, it is known that a delayed coupling can be a source of oscillations, and in some cases it prevents phase locking~\cite{Erneux2009}. However, frequency locking is more general than phase locking, and the oscillatory behavior of coupled, delayed systems is not incompatible with bounded-phase synchronization. Under our experimental conditions, the effect of the delay could not be observed, because the feedback is seen as instantaneous by the laser. Indeed, the typical time scale of the laser dynamics is $ 1/f_R \simeq 10 \, \mu$s, which is much longer than the external-cavity round-trip time $\tau \simeq$ 5 ns. On the contrary, the effect of delay could be conveniently studied using semiconductor lasers, which evolve on much faster time scales~\cite{Sciamanna2007, Sciamanna2009}.

\textit{Acknowledgments.}

We thank F. Bondu for his help on the phase noise calculations. This work has been partly supported by the ``Contrat de Projet Etat R\'egion Bretagne'' PONANT.


\begin{thebibliography}{99}
\bibitem{Strogatz2003}
S. H. Strogatz, \emph{Sync: how order emerges from chaos in the universe, nature and daily life}, (Hyperion, 2003).
\bibitem{Pikovsky2003}
A. Pikovsky, M. Rosenblum, and J. Kurths, \emph{Synchronization: A Universal Concept in Nonlinear Sciences}, (Cambridge University, 2003).
\bibitem{Wiesenfeld1996}
K.Wiesenfeld, P. Colet, and S. H. Strogatz, ``Synchronization transition in a disordered Josephson series array,''
\prl \textbf{76}, 404--407 (1996).
\bibitem{Toiya2010}
M. Toiya, H. O. Gonzalez-Ochoa, V. K. Vanag, S. Fraden, and I. R. Epstein, ``Synchronization of chemical micro-oscillators,'' 
J. Phys. Chem. Lett. \textbf{1}, 1241--1246 (2010).
\bibitem{Fell2011}
J. Fell and J. Axmacher, ``The role of phase synchronization in memory processes,'' 
Nat. Rev. Neurosci. \textbf{12}, 105--118 (2011).
\bibitem{Arenas2008}
A. Arenas, A. Diaz-Guilera, J. Kurths, Y. Moreno, and C. Zhou, ``Synchronization in complex networks,'' 
Phys. Rep. \textbf{469}, 93--153 (2008).
\bibitem{Wieczorek2005}
S. Wieczorek, B. Krauskopf, T. B. Simpson, and D. Lenstra, ``The dynamical complexity of optically injected semiconductor lasers,'' Phys. Rep. \textbf{416}, 1--128 (2005).
\bibitem{Erneux2010} T. Erneux and P. Glorieux, \emph{Laser Dynamics}, (Cambridge University, 2010).
\bibitem{Grillot2009}
N. A. Naderi, M. Pochet, F. Grillot, N. B. Terry, V. Kovanis, and L. F. Lester, ``Modeling the injection-locked behavior of a quantum dash semiconductor laser,'' IEEE J. Sel. Top. Quantum Electron. \textbf{15}, 563--571 (2009).
\bibitem{Hwang2013}
Y. Hung, C. Chu, and S. Hwang, ``Optical double-sideband modulation to single-sideband
modulation conversion using period-one nonlinear
dynamics of semiconductor lasers for radio-over-fiber links,'' \ol \textbf{38}, 1482--1484 (2013).
\bibitem{Kelleher2010}
B. Kelleher, D. Goulding, B. Baselga-Pascual, S. P. Hegarty, and G. Huyet, ``Phasor plots in optical injection experiments,'' Eur. Phys. J. D \textbf{58}, 175--179 (2010).
\bibitem{Thevenin}
J. Th\'evenin, M. Romanelli, M. Vallet, M. Brunel, and T. Erneux, ``Resonance assisted synchronization of coupled oscillators: frequency locking without phase locking,'' \prl \textbf{107}, 104101 (2011).
\bibitem{Braza1990}
P. A. Braza and T. Erneux, ``Constant phase, phase drift, and phase entrainment in lasers with an injected signal,'' \pra \textbf{41}, 6470--6479 (1990).
\bibitem{Solari1994}
H. G. Solari and G.-L. Oppo, ``Laser with injected signal: perturbation of an invariant circle,'' 
Opt. Commun. \textbf{111}, 173--190 (1994).
\bibitem{Strogatz1998}
M. K. S. Yeung and S. H. Strogatz, ``Nonlinear dynamics of a solid-state laser with injection,'' \pre \textbf{58}, 4421--4435 (1998); M. K. S. Yeung and S. H. Strogatz, ``Erratum: Nonlinear dynamics of a solid-state laser with injection,'' \pre \textbf{61}, 2154--2154 (2000).
\bibitem{Ludge2012}
J. Pausch, C. Otto, E. Tylaite, N. Majer,
E. Sch\"oll, and Kathy L\"udge, ``Optically injected quantum dot lasers: impact of
nonlinear carrier lifetimes on frequency-locking
dynamics,''  New Journal of Physics \textbf{14}, 053018 (2012).
\bibitem{Ludge2013} B. Lingnau, W. W. Chow, E. Sch\"oll, and Kathy L\"udge, ``Feedback and injection locking instabilities in quantum-dot lasers: a microscopically based bifurcation analysis,''  New Journal of Physics \textbf{15}, 093031 (2013).
\bibitem{Kronauer1982}
R. E. Kronauer, C. A. Czeisler, S. F. Pilato, M. C. Moore-Ede, and E. D. Weitzman, ``Mathematical model of the human circadian system with two interacting oscillators,'' 
Am. J. Physiol. \textbf{242}, R3-R17 (1982).
\bibitem{Chakraborty1988}
T. Chakraborty and R. H. Rand, ``The transition from phase locking to drift in a system of two weakly coupled van der Pol oscillators,'' Int. J. Non-Linear Mech. \textbf{23}, 369-376 (1988).
\bibitem{Kelleher2012}
B. Kelleher, D. Goulding, B. Baselga Pascual, S. P. Hegarty, and G. Huyet, ``Bounded phase phenomena in the optically injected laser,'' \pre \textbf{85}, 046212 (2012).
\bibitem{Aronson1990}
D. G. Aronson, G. B. Ermentrout, and N. Kopell, ``Amplitude response of coupled oscillators,'' Physica D \textbf{41}, 403--449 (1990).
\bibitem{Thevenin2012}
J. Th\'evenin, M. Romanelli, M. Vallet, M. Brunel, and T. Erneux, ``Phase and intensity dynamics of a two-frequency laser submitted to resonant frequency-shifted feedback,'' 
\pra \textbf{86}, 033815 (2012).
\bibitem{Brunel2004}
M. Brunel, N. D. Lai, M. Vallet, A. Le Floch, F. Bretenaker, L. Morvan, D. Dolfi, J.-P. Huignard, S. Blanc, and T. \nolinebreak Merlet, ``Generation of tunable high-purity microwave and terahertz signals by two-frequency solid state lasers,'' 
Proc. SPIE 5466, Microwave and Terahertz Photonics, 131-139 (2004).
\bibitem{Brunel1997}
M. Brunel, O. Emile, F. Bretenaker, A. Le Floch, B. Ferrand, and E. Molva, 
``Tunable two-frequency lasers for lifetime measurements,'' Optical Review \textbf{4}, 550--552 (1997).
\bibitem{Kervevan}
L. Kervevan, H. Gilles, S. Girard, and M. Laroche, ``Beat-note jitter suppression in a dual-frequency laser using optical feedback,'' \ol \textbf{32}, 1099--1101 (2007).
\bibitem{Rubiola2008}
E. Rubiola, \emph{Phase noise and frequency stability in oscillators}, (Cambridge University, 2008).
\bibitem{IEEE2008}
IEEE Standard Definitions of Physical Quantities for Fundamental Frequency and Time Metrology, IEEE Standard 1139-2008.
\bibitem{Brunel2004b}
M. Brunel, F. Bretenaker, S. Blanc, V. Crozatier, J. Brisset, T. Merlet, and A. Poezevara, ``High-spectral purity RF beat note generated by a two-frequency solid-state laser in a dual thermooptic and electrooptic phase-locked loop,'' IEEE Photon. Technol. Lett. \textbf{16}, 870-872 (2004).
\bibitem{Marquardt2011}
G. Heinrich, M. Ludwig, J. Qian, B. Kubala, and F. Marquardt, ``Collective Dynamics in Optomechanical Arrays,'' \prl \textbf{107}, 043603 (2011).
\bibitem{Agrawal2013}
D. K. Agrawal, J. Woodhouse, and A. A. Seshia, ``Observation of Locked Phase Dynamics and Enhanced Frequency Stability in Synchronized Micromechanical Oscillators,'' \prl \textbf{111}, 084101 (2013).
\bibitem{Soriano} M. C. Soriano, J. Garc\'ia-Ojalvo, C. R. Mirasso, and I. Fischer, ``Complex photonics: Dynamics and applications of delay-coupled semiconductors lasers,'' \rmp \textbf{85}, 421--470 (2013).
\bibitem{Erneux2009} T. Erneux, \emph{Applied delay differential equations}, (Springer, 2009).
\bibitem{Sciamanna2007} M. Sciamanna, I. Gatare, A. Locquet, and K. Panajotov, ``Polarization synchronization in unidirectionally coupled vertical-cavity surface-emitting lasers with orthogonal optical injection,'' \pre \textbf{75}, 056213 (2007).
\bibitem{Sciamanna2009} M. Ozaki, H. Someya, T. Mihara, A. Uchida, S. Yoshimori, K. Panajotov, and M. Sciamanna, ``Leader-laggard relationship of chaos synchronization in mutually coupled vertical-cavity surface-emitting lasers with time delay,'' \pre \textbf{79}, 026210 (2009).
\end{thebibliography}
\end{document}